\documentclass[twocolumn, aps, superscriptaddress]{revtex4-1}

\usepackage{mathptmx}
\usepackage{subfigure}
\usepackage{dcolumn}
\usepackage{amsmath,amssymb}
\usepackage{bm}
\usepackage{color}
\usepackage{latexsym}
\usepackage[english]{babel}
\usepackage{times}
\usepackage{wasysym}

\usepackage{psfrag,graphicx}
\usepackage{epsf}
\usepackage{amsfonts}
\usepackage{natbib}
\usepackage{epstopdf}
\usepackage{appendix}
\usepackage[colorlinks=true,citecolor=blue,linkcolor=blue]{hyperref}

\begin{document}

\title{Excitonic states of an impurity in a Fermi gas}

\author{Zhihao Lan}
\affiliation{Mathematical Sciences, University of Southampton, Highfield, Southampton, SO17 1BJ, UK}
\affiliation{Present address: School of Physics and Astronomy, University of Nottingham, Nottingham, NG7 2RD, UK}
\email{zhihao.lan@nottingham.ac.uk}

\author{Carlos Lobo}
\affiliation{Mathematical Sciences, University of Southampton, Highfield, Southampton, SO17 1BJ, UK}

\begin{abstract}
	
We study excitonic states of an atomic impurity in a Fermi gas, i.e., bound states consisting of the impurity and a hole. Previous studies considered bound states of the impurity with particles from the Fermi sea where the holes only formed part of the particle-hole dressing. Within a two-channel model,  we find that, for a wide range of parameters, excitonic states are not ground but metastable states. We further calculate the decay rates of the excitonic states to polaronic  and dimeronic  states and find they are long lived, scaling as $\Gamma^{\rm{Exc}}_ {\rm{Pol}} \propto ( \Delta\omega)^{5.5}$ and $\Gamma^{\rm{Exc}}_ {\rm{Dim}} \propto (\Delta\omega)^{4}$. We also find that a new continuum of exciton-particle states should be considered alongside the previously known dimeron-hole continuum in spectroscopic measurements. Excitons must therefore be considered as a new ingredient in the study of metastable physics currently being explored experimentally.

\end{abstract}

\maketitle
\section{Introduction}
 An impurity interacting strongly with a background gas is a paradigm of many-body physics \cite{mahan}. This problem has recently attracted great interest due to experimental advances where interactions, mass imbalance and even the dimensionality of the system can be controlled, thus allowing a clean platform to study various properties. Up to now, a handful of states such as the {\em polaron} \cite{lobo1, lobo2}, dressed dimer or {\em dimeron} \cite{Prokofev_Polaron,  Zwerger, MoraChevy, CGL}, and {\em trimeron} \cite{malykh, trimer_parish} have been studied (for a review, see \cite{lan_review, bruun_review}). They can be best understood as quasiparticles: few-body states dressed by particle-hole fluctuations of the Fermi sea whose vacuum properties are qualitatively preserved in the gas.

\begin{figure}
\centering
\includegraphics[width=1.0\columnwidth]{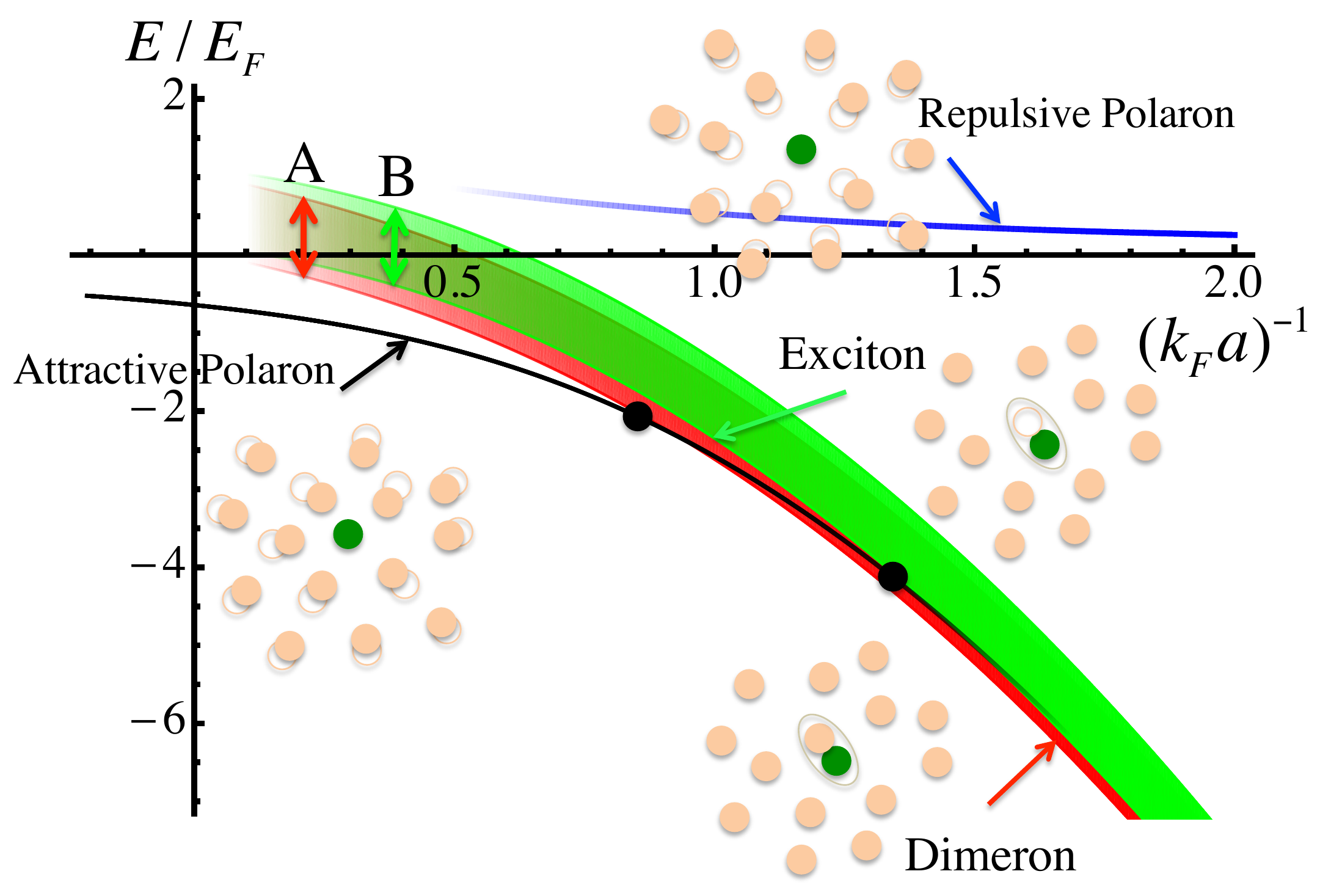}
\caption{ (color online) Schematic for equal masses in the wide resonance limit of the key concepts discussed in the paper. Besides the dimeron-hole continuum identified previously (red band spanned by vectors A) \cite{Grimm}, there is also a metastable exciton-particle continuum (green band spanned by vectors B) and related exciton physics. The bottom of the A band corresponds to the dimeron energy while that of the B band is the exciton energy and the tops are estimated as being roughly the quasiparticle energies plus $\epsilon_F$.}
\label{f1}
\end{figure}

Surprisingly, for the polaron, a simple variational calculation \cite{chevy06} with only one particle-hole dressing gives very accurate results compared with more sophisticated Monte Carlo calculations \cite{Prokofev_Polaron, Houcke3D} and can be extended to study other possible states of the system. For example, away from the unitary regime and approaching the BEC side (Fig. \ref{f1}), the impurity binds one particle from the Fermi sea and forms a dimeron ground state \cite{Zwerger, MoraChevy}. The trimeron becomes the ground state if the ratio of the masses of the impurity to that of the Fermi sea atoms is above a critical mass ratio \cite{trimer_parish}. 

The polaron, dimeron and trimeron have well-defined vacuum limits when the density of the background gas tends to zero (the bare impurity, dimer and trimer) \cite{footnote}. Therefore it is interesting to ask whether there are states which do not have a well-defined vacuum limit. A new such class would be bound states where holes play an essential role (beyond being part of the particle-hole dressing). For example, a bound state of the impurity with a hole could not be interpreted as such in the low density limit but its excitonic-type physics would come into play at nonzero densities. It is interesting to study this possibility in this context; excitonic states are, of course, well known already in charged systems, particularly in solid state  \cite{mahan}.

In this paper, we study variationaly the simplest case of an impurity bound with a single hole (an {\em exciton}) within a two-channel model characterised by a scattering length and effective range in 3D and compare its energy with that of the polaron and dimeron (see Fig.\ref{f1}). We find that the exciton is not the ground state and calculate its decay rates to the polaronic and dimeronic branches, finding that it is a long-lived quasiparticle. An important consequence is that the existence of the exciton gives rise to an exciton-particle continuum in inverse-rf measurements which overlaps with the already known dimeron-hole continuum (\cite{bruun_review,Grimm}) as we see in Fig. \ref{f1}. As a consequence, it is likely that part of the spectral weight of the continuum measured in that experiment and identified as dimeron-hole was most likely of the exciton-particle type instead.

\section{Model}

 We consider an atomic impurity of mass $M$ immersed in a spin polarized Fermi gas of mass $m$. The interaction between the impurity and the background atoms is tunable through an s-wave Feshbach resonance  described by a two-channel model and the interactions between the background atoms can be neglected at low temperatures. The Hamiltonian describing the system is given by (e.g., \cite{trefzger})
\begin{gather}
\hat{H}=\sum_{\mathbf{k}}\left[ \epsilon^u_{\mathbf{k}}\hat{u}_{\mathbf{k}}^{\dagger}\hat{u}_{\mathbf{k}}+\epsilon^d_{\mathbf{k}}\hat{d}_{\mathbf{k}}^{\dagger}\hat{d}_{\mathbf{k}}
+\left(E_{\mathbf{k}}+\nu_0\right)\hat{b}_{\mathbf{k}}^{\dagger}\hat{b}_{\mathbf{k}}\right] 
\nonumber \\
+\frac{g}{\sqrt{V}}\sum_{\mathbf{k,k'}}(\hat{b}_{\mathbf{k+k'}}^{\dagger}\hat{u}_{\mathbf{k'}}\hat{d}_{\mathbf{k}}+\rm{H.c.})
\end{gather}
where the coupling of the impurity and a background atom into the closed channel molecule is described by $g$. Here  $\hat{u}_{\mathbf{k}}^{\dagger}$ ($\hat{d}_{\mathbf{k}}^{\dagger}$ and $\hat{b}_{\mathbf{k}}^{\dagger}$) creates a background atom (an impurity and a closed channel molecule) with wave vector $\mathbf{k}$. The $\hat{u}_{\mathbf{k}}$ obey the usual fermionic anticommutation relations $\{ \hat{u}_{\mathbf{k}}^{\dagger}, \hat{u}_{\mathbf{k'}}\}=\delta_{\mathbf{kk'}}$ and the statistics of the impurity and the molecule are irrelevant since there is at most one impurity and one closed channel molecule in the system. We define $\epsilon^u_{\mathbf{k}} \equiv\hbar^2{\mathbf{k}}^2/2m$, $\epsilon^d_{\mathbf{k}} \equiv \hbar^2{\mathbf{k}}^2/2M$, $E_{\mathbf{k}} \equiv\hbar^2{\mathbf{k}}^2/2(M+m)$ where $\nu_0$ is the internal energy of the molecule. The bare parameters $g$ and $\nu_0$ of the Hamiltonian are related to the physical parameters, scattering length $a$ and effective range of the resonance $R_{*}$ as \cite{trefzger}
\begin{gather}
\frac{\nu_0}{g^2}=-\frac{\mu}{2\pi\hbar^2a}+\int\frac{d^3k}{(2\pi)^3}\frac{2\mu}{\hbar^2k^2}, \hspace{0,5cm}
R_{*}=\frac{\pi\hbar^4}{g^2\mu^2}
\end{gather}
where $\mu=mM/(m+M)$ is the reduced mass. To describe the excitonic state of momentum $|\mathbf{p} \rangle$ of the impurity in the Fermi sea, we use the grand canonical ensemble \cite{lan_review} and consider the variational wave function with one particle-hole dressing,
\begin{gather}
|\Psi_E\rangle_{\mathbf{p}}=\left(\sum'_{\mathbf{q}} \xi_{\mathbf{q}} \hat{d}^{\dagger}_{\mathbf{p+q}} \hat{u}_{\mathbf{q}} +\sum'_{\mathbf{qq'}}\xi_{\mathbf{qq'}} \hat{b}^{\dagger}_{\mathbf{p+q+q'}}  \hat{u}_{\mathbf{q'}} \hat{u}_{\mathbf{q}} \right.  \nonumber \\  +\left. \sum'_{\mathbf{kqq'}}\xi_{\mathbf{kqq'}} \hat{d}^{\dagger}_{\mathbf{p+q+q'-k}} \hat{u}^{\dagger}_{\mathbf{k}}  \hat{u}_{\mathbf{q'}} \hat{u}_{\mathbf{q}}   \right)|\rm{FS}\rangle \label{excitoneq}
\end{gather}
together with the standard ans\"atze for the polaron and dimeron \cite{lan_review, bruun_review}, 
\begin{gather}
|\Psi_P\rangle_{\mathbf{p}}=\left(\phi \hat{d}^{\dagger}_{\mathbf{p}}  +\sum'_{\mathbf{q}}\phi_{\mathbf{q}} \hat{b}^{\dagger}_{\mathbf{p+q}}   \hat{u}_{\mathbf{q}}    + \sum'_{\mathbf{kq}}\phi_{\mathbf{kq}} \hat{d}^{\dagger}_{\mathbf{p+q-k}} \hat{u}^{\dagger}_{\mathbf{k}} \hat{u}_{\mathbf{q}} \right)  |\rm{FS}\rangle \\
|\Psi_D\rangle_{\mathbf{p}}=\left( \eta\hat{b}^{\dagger}_{\mathbf{p}}  + \sum'_{\mathbf{k}} \eta_{\mathbf{k}} \hat{d}^{\dagger}_{\mathbf{p-k}} \hat{u}_{\mathbf{k}}^{\dagger}  +\sum'_{\mathbf{kq}}\eta_{\mathbf{kq}} \hat{b}^{\dagger}_{\mathbf{p+q-k}}  \hat{u}_{\mathbf{k}}^{\dagger} \hat{u}_{\mathbf{q}}  \right. \nonumber \\  + \left. \sum'_{\mathbf{kk'q}}\eta_{\mathbf{kk'q}} \hat{d}^{\dagger}_{\mathbf{p+q-k-k'}} \hat{u}^{\dagger}_{\mathbf{k'}} \hat{u}^{\dagger}_{\mathbf{k}} \hat{u}_{\mathbf{q}}  \right) |\rm{FS}\rangle
\end{gather}
\noindent
where $|\rm{FS}\rangle$ is the zero-temperature noninteracting Fermi sea of background atoms with a Fermi energy $\epsilon_F$ and wave vector $\hbar^2 k_F^2 \equiv 2m \epsilon_F$. Here, $\sum'$  means that the summation of the hole momentum $\mathbf{q}$ is restricted to $q<k_F$ while the particle momentum is restricted to $k>k_F$.  

The equations that determine the polaron, dimeron and exciton energies are given in the large system limit by the minimisation of $\langle \Psi|\hat{H}-\epsilon_F \hat{N} - E|\Psi\rangle$, $\hat{N}$ being the total number operator for background atoms. Setting the zero of the energy at that of the grand canonical energy of a gas of noninteracting atoms with the same $k_F$, we get
\begin{gather}
E_{\rm{pol}}-\epsilon^d_{\mathbf{p}}=\frac{1}{V}\sum'_{\mathbf{q}}\frac{1}{\frac{1}{g^2}E^{\rm{pol}}_{\mathbf{q}}-\frac{1}{V}\sum'_{\mathbf{k}}\frac{1}{E^{\rm{pol}}_{\mathbf{k+q}}}} \\
 \left(\frac{1}{E^{\rm{exc}}_{\mathbf{q}}}\frac{1}{V}\sum'_{\mathbf{q'}}\frac{1}{\Omega^{\rm{exc}}_{\mathbf{qq'}}}-1\right)\chi_{\mathbf{q}}=\frac{1}{V}\sum'_{\mathbf{q'}}\frac{\chi_{\mathbf{q'}}}{E^{\rm{exc}}_{\mathbf{q'}}\Omega^{\rm{exc}}_{\mathbf{qq'}}}  \label{excitonequation} \\
 \Omega^{\rm{dim}}_{\mathbf{kq}}\phi_{\mathbf{kq}}= \frac{1}{V^2}
\sum'_{\mathbf{k'q'}}\frac{\phi_{\mathbf{k'q'}}}{\gamma E^{\rm{dim}}_{\mathbf{k'}} E^{\rm{dim}}_{\mathbf{k}}} +\frac{1}{V}\sum'_{\mathbf{q'}}\frac{\phi_{\mathbf{kq'}}}{E^{\rm{dim}}_{\mathbf{k}}}-\frac{1}{V}\sum'_{\mathbf{k'}}\frac{\phi_{\mathbf{k'q}}}{E^{\rm{dim}}_{\mathbf{kk'q}}} 
\end{gather}
where the various quantities in the above equations are defined as 
\begin{gather}
E^{\rm{pol}}_{\mathbf{q}}=E_{\rm{pol}}-E_{\mathbf{p+q}}-\nu_0+\epsilon^u_{\mathbf{q}} \nonumber \\
E^{\rm{pol}}_{\mathbf{k+q}}=E_{\rm{pol}}-\epsilon^d_{\mathbf{p+q-k}}-\epsilon^u_{\mathbf{k}}+\epsilon^u_{\mathbf{q}},
\end{gather}

\begin{gather}
E^{\rm{exc}}_{\mathbf{q}}=E_{\rm{exc}}-\epsilon^d_{\mathbf{p+q}}+\epsilon^u_{\mathbf{q}}-\epsilon^u_{F} \nonumber \\
E^{\rm{exc}}_{\mathbf{qq'}}=E_{\rm{exc}}-E_{\mathbf{p+q+q'}}-\nu_0+\epsilon^u_{\mathbf{q}}+\epsilon^u_{\mathbf{q'}}-\epsilon^u_{F}\nonumber \\
E^{\rm{exc}}_{\mathbf{kqq'}}=E_{\rm{exc}}-\epsilon^d_{\mathbf{p+q+q'-k}}-\epsilon^u_{\mathbf{k}}+\epsilon^u_{\mathbf{q}}+\epsilon^u_{\mathbf{q'}}-\epsilon^u_{F}\nonumber \\
\Omega^{\rm{exc}}_{\mathbf{qq'}}=\frac{E^{\rm{exc}}_{\mathbf{qq'}}}{g^2}-\frac{1}{V}\sum'_{\mathbf{k}}\frac{1}{E^{\rm{exc}}_{\mathbf{kqq'}}}, \hspace{0.4cm} \chi_{\mathbf{q}}=\frac{1}{V}\sum'_{\mathbf{q'}}\xi_{\mathbf{qq'}}, \end{gather}
and 
\begin{gather}
E^{\rm{dim}}_{\mathbf{k}}=E_{\rm{dim}}-\epsilon^d_{\mathbf{p-k}}-\epsilon^u_{\mathbf{k}}+\epsilon_F\nonumber \\
E^{\rm{dim}}_{\mathbf{kq}}=E_{\rm{dim}}-E_{\mathbf{p+q-k}}-\nu_0-\epsilon^u_{\mathbf{k}}+\epsilon^u_{\mathbf{q}}+\epsilon_F \nonumber \\
E^{\rm{dim}}_{\mathbf{kk'q}}=E_{\rm{dim}}-\epsilon^d_{\mathbf{p+q-k-k'}}-\epsilon^u_{\mathbf{k'}}-\epsilon^u_{\mathbf{k}}+\epsilon^u_{\mathbf{q}}+\epsilon_F\nonumber \\
\Omega^{\rm{dim}}_{\mathbf{kq}}=\frac{E^{\rm{dim}}_{\mathbf{kq}}}{g^2}- \frac{1}{V}\sum'_{\mathbf{k'}} \frac{1}{E^{\rm{dim}}_{\mathbf{kk'q}}}\nonumber \\
\gamma=(E_{\rm{dim}}-E_{\mathbf{p}}-\nu_0+\epsilon_F)/g^2-\frac{1}{V}\sum'_{\mathbf{k}}\frac{1}{E_{\mathbf{k}}},
\end{gather}
\noindent
for polaron, exciton and dimeron respectively. Note that the energy equation of polaron is algebraic while that for the exciton and dimeron are integral equations.  From the above equations, we can extract the exciton, polaron and dimeron energies as a function of various parameters, like, $a$, $R_{*}$ and $M/m$ (see \cite{Zwerger, trefzger}). In the next section we will consider the cases with $\mathbf{p}=0$.

\begin{figure}
\centering
\includegraphics[width=1.0\columnwidth]{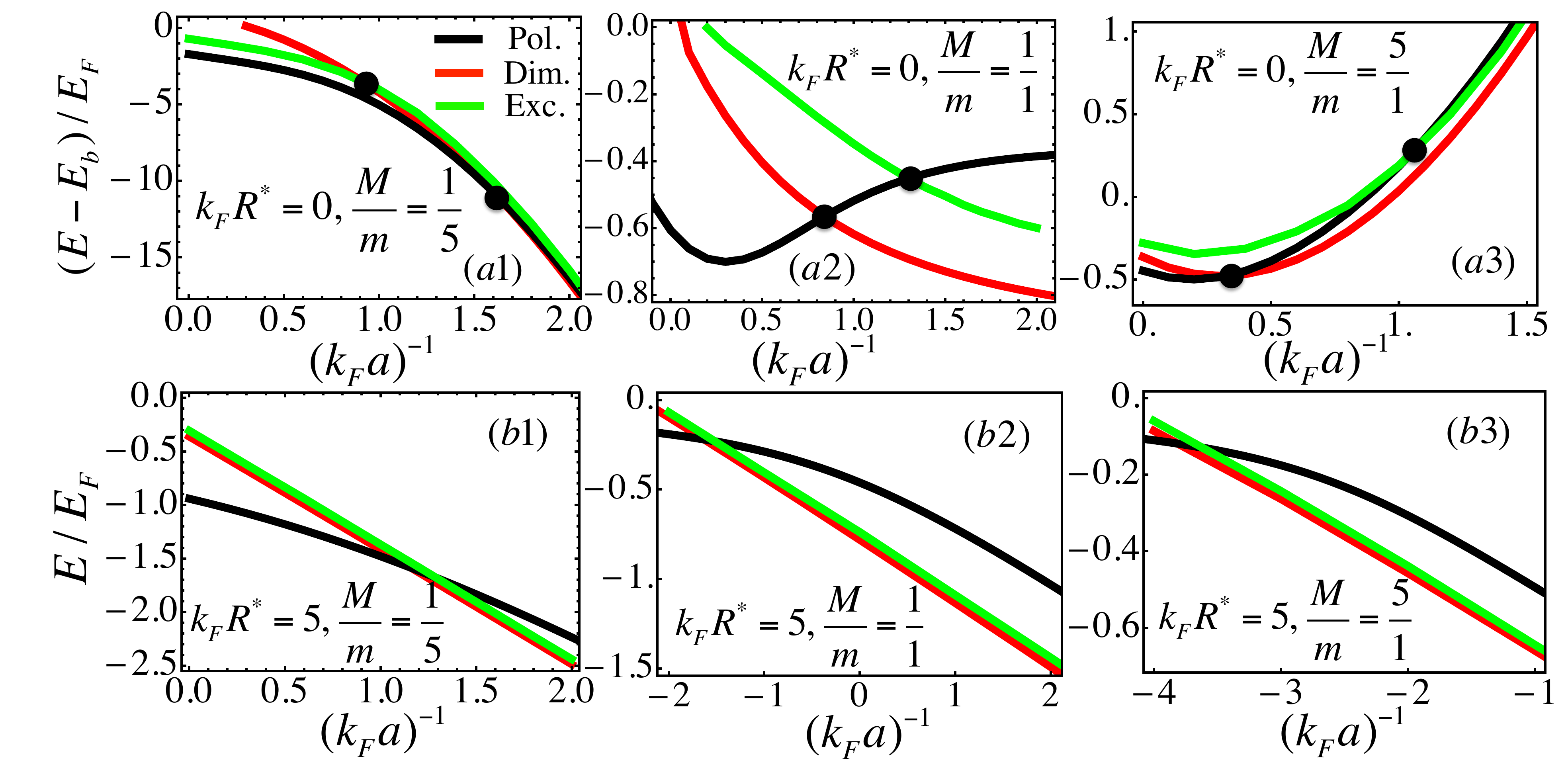}
\caption{ (color online) The energies of the polaron, exciton and dimeron as a function of the interaction $(k_Fa)^{-1}$ with different mass $M/m$ and resonance parameter $kR^*$. Panels (a1)-(a3) show the energies when $k_F R^*=0$ at three different masses $M/m=1/5,1,5$ where the binding energy $E_b=-\hbar^2/ma^2$ has been subtracted for clarity. The dots mark the crossing points. Panels (b1)-(b3) show the energies when $k_F R^*=5$. }
\label{f2}
\end{figure}

\section{Energy of the exciton}
 In Fig. \ref{f2} we present the energies of the exciton compared with that of the polaron and dimeron as a function of $(k_Fa)^{-1}$ at different  $M/m$ and $k_FR^*$.   Panel a(2) of Fig.\ref{f2} shows the calculations with equal mass ($M/m=1$) and in the wide resonance limit ($k_FR^*=0$), where the energies of the polaron and dimeron have been studied previously \cite{Zwerger, MoraChevy, CGL, trefzger, Houcke3D}. The energies of the polaron and dimeron cross at $(k_Fa_c)^{-1}\simeq 0.85$ while that of  the exciton and polaron cross at $(k_Fa_c)^{-1}\simeq 1.3$. Increasing the mass ratio $M/m$, as in Fig.\ref{f2} (a3) where $M/m=5$, the critical values of $(k_Fa_c)^{-1}$ for polaron-dimeron and exciton-polaron crossings both decrease, but the exciton energy is alway higher than that of the dimeron. The situation changes when the mass of the impurity is lighter than that of the atoms of the Fermi sea, as in Fig.\ref{f2} (a1) where $M/m=1/5$ and the exciton and dimeron energies cross at $(k_Fa_c)^{-1}\simeq 0.9$, though the polaron energy is lower than that of the exciton at this scale. Fig. \ref{f2} (b1)-(b3) shows the energies of the three states when the resonance becomes narrow ($k_FR^*=5$). In this case, the energy curves of the dimeron and exciton become essentially flat around unitarity and the exciton energy is alway higher than that of the dimeron for all three mass ratios of $M/m=1/5, 1, 5$. From this we see that the exciton is never the ground state but is a resonance whose lifetime we discuss below.

Because we are studying variationaly a quasiparticle which is not the lowest energy state (as was the case of the repulsive polaron in \cite{cui_zhai, 2d_meera2}), we should check that ansatz (\ref{excitoneq}) is indeed describing a new quasiparticle and not some trivial combination of other known ones which are also included in the Hilbert space spanned by the ansatz. For example, it might describe a polaron plus a hole or (if the first term in (\ref{excitoneq}) were zero) a dimeron plus two holes. The energy of the latter two states is approximately given by the sum of the energies of the quasiparticles (e.g. polaron energy plus hole energy) with the constraint of zero total momentum. We argue as follows: if we look at panels (a1) and (a2) of Fig. \ref{f2}, we see that, in the first case, the exciton has lower energy than the dimeron for values smaller than $(k_Fa)^{-1} \simeq 0.9$. Therefore its state for $M/m=1/5$ and for all $k_Fa$ is not trivial combination of the dimeron plus two holes which is of equal or higher energy. In the central panel, we can similarly argue that the exciton is lower in energy than the state of a polaron plus a hole for $M/m=1$ since its energy is lower than that of a polaron for $(k_Fa)^{-1}>1.3$. It follows then that, in the $(M/m,(k_Fa)^{-1})$ plane, the exciton eigenstate is not trivial  combination of those two states everywhere by continuity. Clearly, it will also not be trivial  combination of other states that are included in the Hilbert space spanned by (\ref{excitoneq}) such as a ``Cooper pair" (dimer without dressing) plus two holes or single impurity plus hole since these are higher in energy than the dimeron or polaron respectively. From this we conclude that the exciton is indeed a new quasiparticle.

\section{Diagrammatic formulation and lifetime of the exciton}

\begin{figure}
\centering
\includegraphics[width=1.0\columnwidth]{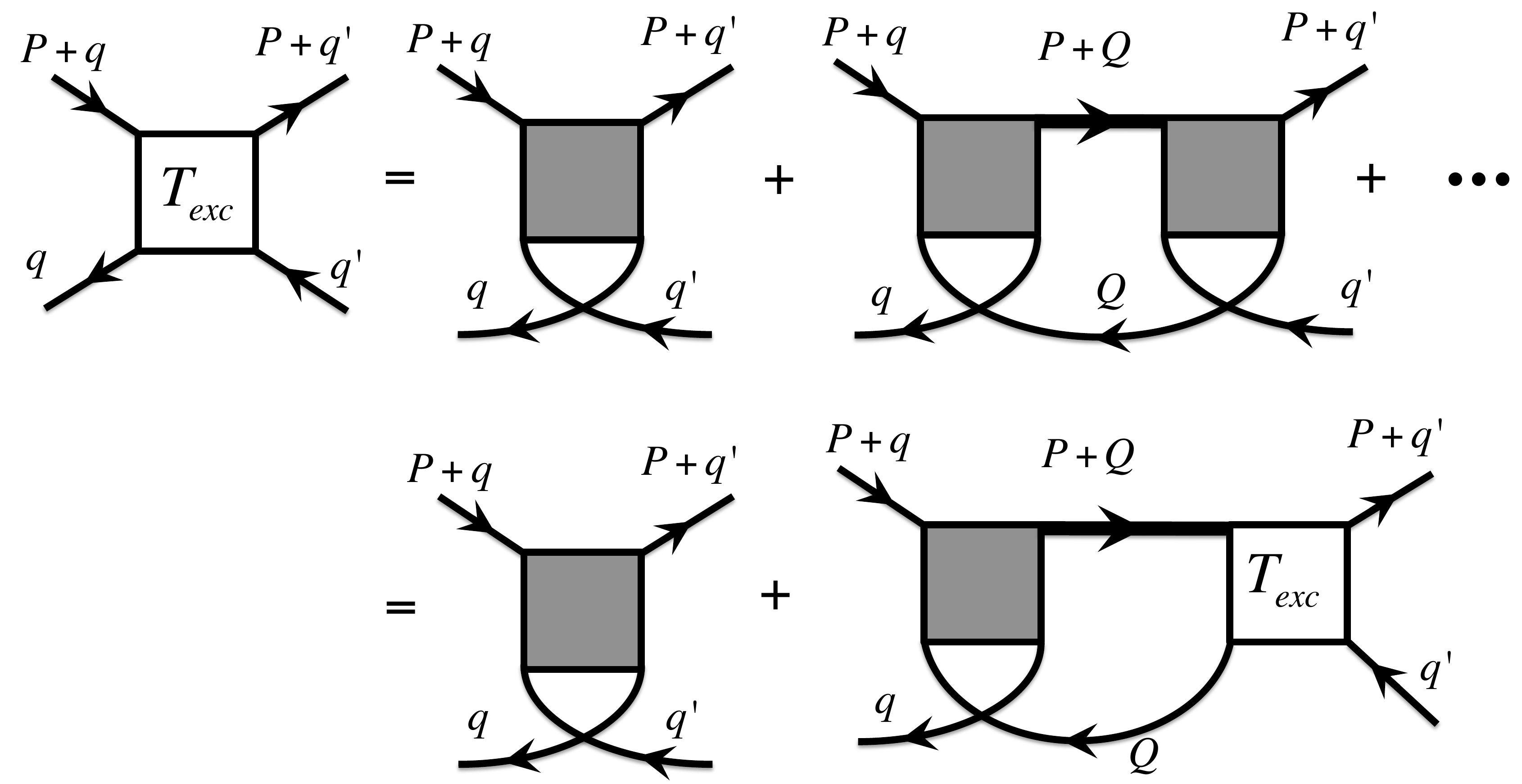}
\caption{Equation for the T-matrix of the scattering of an impurity with a hole whose resonance is the exciton. Right moving lines are impurity lines and left moving are background atoms (holes). The bold line is a polaron dressing (up to one particle-hole) and the grey squares are the usual impurity-background atom T-matrix in the ladder approximation, see also Fig. (\ref{f4}).}
\label{f3}
\end{figure}

 The variational approach is strictly speaking valid for eigenstates of the system, usually the ground state. The exciton however is not the ground state but a resonance so, while the variational method can be applied in a restricted Hilbert space as we have argued, it might be useful to present its diagrammatic formulation. The resonance appears in the impurity-hole T-matrix at positive energy and total momentum $\mathbf{p}$ as in (\ref{excitoneq}). In Fig. \ref{f3} we show the diagrams corresponding to ansatz (\ref{excitoneq}), i.e. including at most two particle-hole lines at any given time. We also see that intermediate states correspond to the repeated scattering of a polaron and a hole. These diagrams represent all possible intermediate states which span the two-particle hole Hilbert space of the variational ansatz and are therefore equivalent to it. They also illustrate that the exciton is a true resonance, not just a combination of a polaron and a hole since it requires interaction between the two.  A similar type of diagrams were considered in \cite{Demler}. However they were interested in the equal spin population case and did not consider the polaronic dressing of the impurity lines.

 Since the exciton is not the ground state of the system, the excitonic states can be a long lived quasiparticle at best since they can decay to a dimeron or polaron. If the lifetime is long enough, then we will be able to speak of a new quasiparticle, the exciton, opening a new regime of metastable physics (similarly to the repulsive polaron which is also not the ground state but can still be observed experimentally due to its sufficiently long lifetime and has actually already been studied experimentally by several groups recently \cite{Zwierlein, SalomonScience, Grimm, kohl}). Thus the exciton is a resonance in the scattering of the impurity with a hole, which appears as a pole at the complex energy $E+i\Gamma$ in the scattering amplitude. The real part of the pole corresponds approximately to the exciton energy calculated by the variational ansatz. The important question then becomes - what is the imaginary part of the pole, i.e.,  what is the lifetime of the exciton? As discussed above, these processes require a larger Hilbert space than that of the variational ansatz.

\begin{figure*}
\centering
\includegraphics[width=2.0\columnwidth]{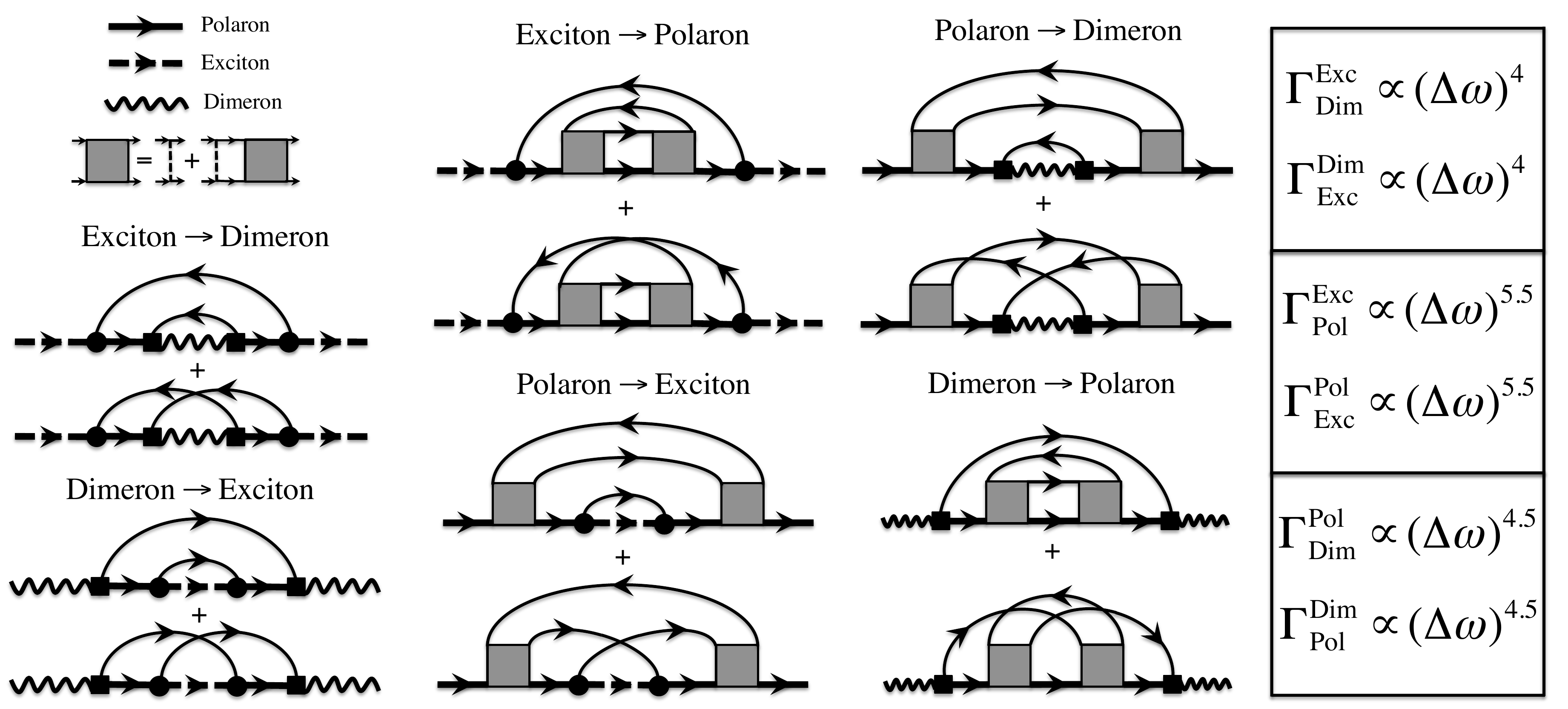}
\caption{The decay processes among the three states: polaron, dimeron and exciton, which are described by the single-particle, two-particle and particle-hole propagators \cite{mattock}. The first (second, third) column shows the diagrams of the decay processes between exciton and dimeron (exciton and polaron, polaron and dimeron). The last column summarises the scaling laws for these decay processes where $\Gamma^{\alpha}_{\beta}$ denotes the decay rate of ``$\alpha$" to ``$\beta$". The thick solid (dashed, wavy)  line is the polaron (exciton and dimeron) propagator while the thin solid line is the propagator of the Fermi background atom. $\CIRCLE$ denotes the exciton-polaron coupling strength $g_1$,  $\blacksquare$ the polaron-dimeron coupling strength $g_2$ and grey square the usual impurity-background atom T-matrix in the ladder approximation.}
\label{f4}
\end{figure*}

Fig. \ref{f2} shows that all three quasiparticles can decay into each other. The Feynman diagrams for these processes are shown in Fig. \ref{f4}. The decay rate $\Gamma^{\alpha}_{\beta}$ (decay from ``$\alpha$" to ``$\beta$") is determined by the imaginary part of the corresponding self energy insertions $\Sigma$ (the part of the diagram without external legs), i.e., $\Gamma=-\rm{Im}\Sigma(0,\omega)$ \cite{fetter}. While the decay rates between polaron and dimeron have been calculated at zero \cite{decay_bruun} and finite momentum \cite{decay_lobo} we will focus on the decay rates of the exciton to the dimeron and polaron. However, for completeness in Fig. \ref{f4} we also show the diagrams for the decay between the polaron and dimeron. 

We first calculate the decay rates from exciton to dimeron, in which process a zero momentum exciton decays to a dimeron and two holes so that the decay rate involves a sum over the hole final momenta. The frequency sums in the two diagrams can be performed by contour integration which amounts to evaluating the frequencies at the on-shell energies \cite{decay_bruun}. Using a pole expansion of the dimeron and polaron propagator, $D(\mathbf{q},\omega)\simeq Z_D/(\omega-\omega_D-q^2/2m^*_D) $ and  $G^d(\mathbf{q},\omega)\simeq Z_P/(\omega-\omega_p-q^2/2m^*_P) $, with $Z_P, Z_D$ and $m_P^*, m_D^*$ the corresponding residues and effective masses, we find (from the imaginary part of the two self-energy insertions \cite{fetter}), 
\begin{gather}
\Gamma^{\rm{Exc}}_{\rm{Dim}}=\frac{\pi g_1^2g_2^2Z_P^2Z_D}{2}\int \frac{d^3\mathbf{q}}{(2\pi)^3}\frac{d^3\mathbf{q'}}{(2\pi)^3} \left[F(q)-F(q')\right]^2\nonumber \\
\delta\left(\Delta \omega_D^E+\xi_q^u+\xi_{q'}^u-\frac{(\mathbf{q+q'})^2}{2m_D^*}\right) 
\sim (\Delta \omega_D^E)^4
\label{emDecay}
\end{gather}  \noindent where $F(q)=G^d(\mathbf{q},\omega_{\rm{Exc}}+\xi_q^u)/Z_P$  and $\Delta \omega_D^E=\omega_{\rm{Exc}}-\omega_{\rm{Dim}}$. It is interesting to note that without the matrix element term $\left[F(q)-F(q')\right]^2$, the decay rate scales as $\sim (\Delta \omega)^2$, so the Fermi antisymmetry when swapping the two hole momenta $\mathbf{q}$ and $\mathbf{q'}$ gives an additional factor $(\Delta \omega)^2$ to the decay rate. This is very different from the case of decay rate between polaron and dimeron where the Fermi antisymmetry only gives   $\Delta \omega$ \cite{decay_bruun} due to the absence of an additional particle from the Fermi sea as a result of the decay.  From a similar calculation for the decay of the dimeron to exciton, we find $\Gamma^{\rm{Dim}}_{\rm{Exc}} \sim (\Delta \omega_{\rm{E}}^{\rm{D}})^4$.

We then proceed to the calculations of the decay rates from exciton to polaron, in which a zero momentum exciton decays into a polaron, a particle and two holes (the energy threshold for decay into a polaron and hole is higher). The two diagrams in the second column of Fig. \ref{f4}  yield the following expression, 
\begin{gather}
\Gamma^{\rm{Exc}}_{\rm{Pol}}=\frac{\pi g_1^2Z_P^3}{2}\int \frac{d^3\mathbf{q}}{(2\pi)^3} \frac{d^3\mathbf{q'}}{(2\pi)^3} \frac{d^3\mathbf{k}}{(2\pi)^3}\left[F_1(\mathbf{q,q'})-F_2(\mathbf{q,q'})  \right]^2 \nonumber \\
\hspace{-0.1cm}\delta\left(\Delta \omega_P^E+\xi_q^u+\xi_{q'}^u-\xi_{k}^u-\frac{(\mathbf{q+q'-k})^2}{2m_P^*}\right) 
\sim (\Delta \omega_P^E)^{5.5}
\label{epDecay}
\end{gather}
\noindent
where $F_1(\mathbf{q,q'})=G^d(\mathbf{q},\omega_{\rm{Exc}}+\xi_q^u)T_2(\mathbf{q}+\mathbf{q'},\omega_{\rm{Exc}}+\xi_q^u+\xi_{q'}^u)/Z_P$, $F_2(\mathbf{q,q'})=G^d(\mathbf{q'},\omega_{\rm{Exc}}+\xi_{q'}^u)T_2(\mathbf{q}+\mathbf{q'},\omega_{\rm{Exc}}+\xi_q^u+\xi_{q'}^u)/Z_P$ and $\Delta \omega_P^E=\omega_{\rm{Exc}}-\omega_{\rm{Pol}}$. Again, we find without the matrix element term the above expression would give a scaling $\sim (\Delta \omega)^{3.5}$, so the Fermi antisymmetry when swapping the two hole momenta $\mathbf{q}$ and $\mathbf{q'}$ also gives an additional factor $(\Delta \omega)^2$ to the decay rate.  This may be a unique feature of the exciton physics in this  setup. A similar analysis of the two diagrams of the polaron to exciton decay gives  $\Gamma^{\rm{Pol}}_{\rm{Exc}} \sim (\Delta \omega^P_E)^{5.5}$.

The scalings of the decay rates among the three states are summarised in the last column of Fig. . These scaling laws with high powers are interesting theoretical results on their own and imply very long quasiparticle lifetimes as compared with the usual quadratic scaling $(\Delta \omega)^{2}$ of the quasiparticle decay rates of a normal Landau fermi liquid (for the physics of low power laws, such as $\Delta \omega$ or even frequency independent decay rates, see \cite{lan_prl}), and will open a new regime of metastable physics in Fermi gases for experimental explorations \cite{decay_lobo}. Relating our scaling laws to experiments, we can estimate the exciton lifetime, for example at the exciton-polaron crossing point $1/k_Fa\simeq 1.3$ of Fig. \ref{f2} (a2). At the crossing point,  the decay of exciton to polaron will vanish, though the exciton can decay to dimeron with an energy difference of $\Delta \omega/\epsilon_F \simeq 0.25$. Using $g_2=-\sqrt{2\pi/\mu^2a}$ \cite{decay_bruun} and assuming $g_1\simeq g_2$, we find the prefactor of the scaling law by doing the integral (\ref{emDecay}) numerically, resulting in $\hbar \Gamma\sim2.4\epsilon_F (\Delta \omega/\epsilon_F)^4\sim0.01\epsilon_F$ (for a Fermi energy of $\epsilon_F=h\times 37kHz$, the lifetime is $\sim0.4ms$), which is the typical scale of the repulsive polaron as measured in \cite{Grimm}.  Relating this decay rate to the corresponding energy shift of the excitonic state ($|E_{-}| \sim0.45 \epsilon_F$), we find $\hbar \Gamma/|E_{-}| \approx 0.02 \ll1$, which shows that the excitonic state exists as a well defined, metastable quasiparticle in this case.

\section{Discussion}

While the excitonic state is not the ground state at least for some parameter values, its long lifetime adds new ingredients to the metastable physics previously known (e.g., repulsive polaron). An interesting result obtained here is that the Fermi antisymmetry when swapping the two hole momenta $\mathbf{q}$ and $\mathbf{q'}$ gives an additional factor $(\Delta \omega)^2$ to the exciton decay rates in contrast with the previously studied decay rates between polaron and dimeron where only an additional factor $(\Delta \omega)$ was found.

Our study opens many interesting directions since it emphasises hole excitations as an integral part of few-body states in the gas. It naturally leads to questions about states with more than a single hole, and to analogies with solid state systems even though in the latter, the physics has qualitatively different aspects due to the quasiparticle charge.  An interesting question is whether the exciton can be the ground state in closely related systems. For example, in 2D \cite{2d_recati, 2d_meera1,2d_schmidt, 2d_meera2, 2d_zollner}, integrals over hole momenta are more important; also, one could think of other states such as p-wave polarons and dimerons \cite{p_polaron} or dipolar impurity systems \cite{dipole_impurity} and even impurities with spin-orbit couplings \cite{SO_polaron}. Further studies will also be needed to clarify how to couple the impurity with the excitonic states while having negligible coupling with the dimeronic or polaronic states by inverse rf (radio-frequency) process which will allow us to probe the excitonic states directly. 

{\it Acknowledgments---}. We acknowledge support from the EPSRC through grant EP/I018514/1. 

%%%%%%%%%%%%%%%%%%%%


\begin{references}
\bibitem{mahan} G. D. Mahan, {\it Many-Particle Physics} (Kluwer Academic,
New York, 2000).
\bibitem{lobo1} C. Lobo, A. Recati, S. Giorgini, and S. Stringari, Phys. Rev. Lett. {\bf 97}, 200403 (2006).
\bibitem{lobo2} R. Combescot, A. Recati, C. Lobo, and F. Chevy, Phys. Rev. Lett. {\bf 98}, 180402 (2007).
\bibitem{Prokofev_Polaron} N. V. Prokofev and B. V. Svistunov,  Phys. Rev. B {\bf 77}, 020408(R) (2008); Phys. Rev. B {\bf 77}, 125101 (2008).
\bibitem{Zwerger} M. Punk, P. T. Dumitrescu, and W. Zwerger, Phys. Rev. A {\bf 80}, 053605 (2009).
\bibitem{MoraChevy} C. Mora and F. Chevy,  Phys. Rev. A {\bf 80}, 033607(2009).
\bibitem{CGL} R. Combescot, S. Giraud,  and X. Leyronas, EPL {\bf 88}, 60007 (2009).
\bibitem{malykh}O. I. Kartavtsev and A. V. Malykh, J. Phys. B: At. Mol. Opt. Phys. {\bf 40}, 1429 (2007).
\bibitem{trimer_parish} C. J. M. Mathy, M. M. Parish, and D. A. Huse, Phys. Rev. Lett. {\bf 106}, 166404 (2011).

\bibitem{lan_review} Z. Lan and C. Lobo, J. Indian I. Sci. {\bf 94}, 179 (2014).
\bibitem{bruun_review} P. Massignan, M. Zaccanti, and G. M. Bruun, Rep. Prog. Phys. {\bf 77}, 034401 (2014).

\bibitem{chevy06} F. Chevy, Phys. Rev. A {\bf 74}, 063628 (2006).
\bibitem{Houcke3D} J. Vlietinck, J. Ryckebusch, and K. VanHoucke, Phys. Rev. B {\bf 87}, 115133 (2013).

\bibitem{footnote} With a minor exception known theoretically \cite{trimer_parish}: a trimer consisting of two heavy background atoms plus one light impurity can be stable in the Fermi gas at mass ratios which make it unstable in vacuum.



\bibitem{Grimm} C. Kohstall, M. Zaccanti, M. Jag, A. Trenkwalder, P. Massignan, G. M. Bruun, F. Schreck, and R. Grimm,  Nature (London) {\bf 485}, 615 (2012).

\bibitem{trefzger} C. Trefzger and Y. Castin, Phys. Rev. A {\bf 85}, 053612 (2012).


\bibitem{cui_zhai} X. Cui and H. Zhai, Phys. Rev. A {\bf 81}, 041602 (R)  (2010).
\bibitem{2d_meera2}M. M. Parish and J. Levinsen, Phys. Rev. A {\bf 87}, 033616 (2013).
\bibitem{Demler} David Pekker, Mehrtash Babadi, Rajdeep Sensarma, Nikolaj Zinner, Lode Pollet, Martin W. Zwierlein, and Eugene Demler
Phys. Rev. Lett. {\bf 106}, 050402 (2011).

\bibitem{Zwierlein} A. Schirotzek, C.-H. Wu, A. Sommer, and M. W. Zwierlein,  Phys. Rev. Lett. {\bf 102}, 230402 (2009). 
\bibitem{SalomonScience}  N. Navon, S. Nascimbene, F. Chevy, and C. Salomon,  Science  {\bf 328}, 729 (2010). 
\bibitem{kohl}M. Koschorreck, D. Pertot, E. Vogt, B.Fr\"{o}hlich, M. Feld, and M. K\"{o}hl, Nature (London) {\bf 485}, 619 (2012).

\bibitem{fetter} A. L. Fetter and J. D. Walecka, {\it Quantum Theory of Many-Particle Systems}, Dover Publications Inc. (2003).
\bibitem{decay_bruun} G. M. Bruun and P. Massignan, Phys. Rev. Lett. {\bf105}, 020403 (2010).
\bibitem{decay_lobo}K. Sadeghzadeh, G. M. Bruun, C. Lobo, P. Massignan, and A. Recati, New J. phys. {\bf 13}, 055011 (2011).

\bibitem{mattock} R. D. Mattuck, {\it A guide to Feynman diagrams in the many-body problem}, Dover Publications Inc. (1992).


\bibitem{lan_prl} Z. Lan, G. M. Bruun, and C. Lobo, Phys. Rev. Lett. {\bf 111}, 145301 (2013).
\bibitem{2d_zollner} S. Z\"{o}llner, G. M. Bruun, and C. J. Pethick, Phys. Rev. A {\bf 83}, 021603 (2011).
\bibitem{2d_meera1} M. M. Parish, Phys. Rev. A {\bf 83}, 051603(R) (2011).
\bibitem{2d_recati} M. Klawunn and A. Recati, Phys. Rev. A {\bf 84}, 033607 (2011).
\bibitem{2d_schmidt}R. Schmidt, T. Enss, V. Pietila, and E. Demler, Phys. Rev. A {\bf 85}, 021602(R) (2012).
\bibitem{p_polaron} J. Levinsen, P. Massignan, F. Chevy, and C. Lobo, Phys. Rev. Lett. {\bf 109}, 075302 (2012).
\bibitem{dipole_impurity} N. Matveeva and S. Giorgini, Phys. Rev. Lett. {\bf111}, 220405 (2013).
\bibitem{SO_polaron}W. Yi and W. Zhang, Phys. Rev. Lett. {\bf 109}, 140402 (2012); L. Zhou, X. Cui, and W. Yi, Phys. Rev. Lett. {\bf 112}, 195301 (2014).

%%%%%%%%%%%%%%%

\end{references}
\end{document}